\documentstyle[12pt]{article}
\newlength{\extraspace}
\newlength{\extraspaces}
\newcommand{\be}{\begin{equation}
\addtolength{\abovedisplayskip}{\extraspaces}
\addtolength{\belowdisplayskip}{\extraspaces}
\addtolength{\abovedisplayshortskip}{\extraspace}
\addtolength{\belowdisplayshortskip}{\extraspace}}
\newcommand{\ee}{\end{equation}}
\newcommand{\ba}{\begin{eqnarray}
\addtolength{\abovedisplayskip}{\extraspaces}
\addtolength{\belowdisplayskip}{\extraspaces}
\addtolength{\abovedisplayshortskip}{\extraspace}
\addtolength{\belowdisplayshortskip}{\extraspace}}
\newcommand{\ea}{\end{eqnarray}}
\begin{document}
\thispagestyle{empty}
\begin{flushright}
SIT-LP/08/10-2 \\
October, 2008
\end{flushright}
\vspace{7mm}
\begin{center}
 {\sc{\large{\bf SUPERSYMMETRY, GENERAL RELATIVITY \\[2mm]
 AND \\[2mm]
 UNITY OF NATURE}}} 
\footnote{
\tt Based on the talk given by K. Shima at EU RTN Workshop 
{\it Constituents, Fundamental Forces and Symmetries of the Universe},
 11-17 September 2008, Varna, Bulgaria.} 
\\[20mm]
{\sc Kazunari Shima and Motomu Tsuda}
\footnote{
\tt e-mail: shima@sit.ac.jp, tsuda@sit.ac.jp
}
\\[5mm]
{\it Laboratory of Physics, 
Saitama Institute of Technology \\
Fukaya, Saitama 369-0293, Japan} \\[20mm]
\begin{abstract}
The basic idea and some physical implications of nonlinear supersymmetric 
general relativity (NLSUSY GR) are presented. 
NLSUSY GR may give new insights into the origin of mass and the mysterious relations between the cosmology 
and the low energy particle physics, e.g. 
the spontaneous SUSY breaking scale, the cosmological constant, 
the (dark) energy density of the universe and the neutrino mass. 
%
\end{abstract}
\end{center}

\newpage

\noindent

\section{Introduction}
The symmetry and its spontaneous breaking are important notions for the unified description of nature.   
Supersymmetry (SUSY) \cite{WZ1,VA,GL} is a profound notion
related naturally to space-time symmetry, which is  promissing for the unification of 
graviton with all other particles in the single irreducible representation 
of the symmetry group. 
Therefore, the evidences of SUSY and its spontaneous breakdown \cite{SS,FI,O} should be studied  
not only in (low energy) particle physics but also in cosmology, 
i.e. in the framework necessarily  accomodating graviton \cite{MG}. \par
Along these viewpoints, we have found by the  group theoretical arguments that 
among all $SO(N)$ super-Poincar\'e (SP) groups  the $SO(10)$ SP group with the decomposition of $10$ supercharges as 
${\underline{10} = \underline{5}+\underline{5^{*}}}$ under $SO(10) \supset SU(5) \supset SU(3) \times SU(2) \times U(1) $
may be a ${\it unique}$ and ${\it minimal}$ group which accomodates all observed particles, 
i.e. the standard model (SM) with just three generations of quarks and leptons including graviton 
are accomodated as SUSY eigenstates in a single irreducible representation of $SO(10)$ SP group, 
which constitute a single $SO(10)$ linear (L) SUSY supermultiplet \cite{KS1}. 
(We remember that the relativistic hydrogen atom is solved by $O(4)$ space-time symmetry.) 
Remarkably we have assigned ${\underline 5}$  the same quantum numbers as those of ${\underline 5}$ of SU(5) GUT 
\cite{GG}, i.e. 
\begin{equation}
{\underline 5} = ( {}\underline 3, \underline 1; [ {1 \over 3}.  {1 \over 3},  {1 \over 3}] )+
( {\underline 1},  {\underline 2}; [ -1, 0 ]) 
\label{qn}
\end{equation}%
with respect to $(\underline {SU(3)}, {\underline {SU(2)}}; [ Q_{e}])$.
Furthermore considering that the superchages (generators) in the SUSY algebra for the massless representation  
of $SO(10)$ SP  in the light-cone frame 
can be interpreted as the creation and annihilation operators for spin $1 \over 2$ particle,
we are tempted to imagine that there may be a certain composite structure (far) beyond the SM.
We have proposed the superon-quintet (SQ) hypothesis \cite{KS3} corresponding to  ${\underline 5}$   
as the fundamental spin ${1\over 2}$ massless objects for all observed particles.  \par
The advocated difficulty for constructing non-trivial $SO(N > 8)$ SUSY (gravity) theory, 
the so called no-go theorem based on S-matrix argument \cite{CM,HLS}, 
can be circumvented by adopting the {\it nonlinear (NL)} representation of SUSY \cite{WB}, 
i.e. the vacuum degeneracy of the fundamental action. 
The NL representation of SUSY gives the (unique) action describing the spontaneous breakdown of SUSY. 
Volkov-Akulov (VA) model \cite{VA} gives the NL representation of $N = 1$ SUSY 
describing the dynamics of spin ${1 \over 2}$ 
Nambu-Goldstone (NG) fermion accompanying the spontaneous SUSY breaking in flat space-time.     \par
Therefore the NLSUSY invariant generalization of the general relativity  
gives the fundamental theory of everything in our scenario, 
i.e. the ultimate shape of nature is unstable empty space-time described by 
the principle of the general relativity \cite{YANG85}.
\section{Nonlinear Supersymmetric General Relativity}
For simplicity we discuss $N=1$ without the loss of the generality. 
The extension to $N>1$ is crucial for the realistic model building.  \par
The fundamental action 
(called nonlinear supersymmetric general relativity (NLSUSY GR)) 
has been constructed by extending the geometric arguments 
of Einstein general relativity (EGR) on Riemann space-time to new space-time inspired by NLSUSY, 
where tangent space-time is specified  not only by the Minkowski coodinate $x_a$ of $SO(1,3)$ 
but also by the Grassmanian coordinate $\psi_\alpha$ of the isomorphic $SL(2C)$ for NLSUSY \cite{KS2,KS3}, 
i.e. the coset parameters of ${super GL(4R) \over GL(4R)}$  
interpreted as NG fermions associated with the spontaneous breaking of super-$GL(4R)$ down to $GL(4R)$. 
(The  noncompact isomorphic groups $SO(1,3)$ and $SL(2C)$ for tangent space-time symmetry on curved space-time 
can be regarded as the generalization of the compact isomorphic groups $SU(2)$ and $SO(3)$ for the gauge symmetry 
of 't Hooft-Polyakov monopole on flat space-time.)  \par
The NLSUSY GR action \cite{KS2,KS3} is given by 
\begin{eqnarray}
& & L_{\rm NLSUSYGR}(w) = {c^4 \over 16{\pi}G} \vert w \vert \{\Omega(w) - \Lambda \}, 
\label{SGM}
\\[2mm]
& & \hspace*{7mm} 
\vert w \vert = \det w^a{}_\mu = \det \{e^a{}_\mu + t^a{}_\mu(\psi)\}, 
\nonumber \\[.5mm]
& & \hspace*{7mm} 
t^a{}_\mu(\psi) = {\kappa^2 \over 2i}(\bar\psi \gamma^a \partial_\mu \psi 
- \partial_\mu \bar\psi \gamma^a \psi), 
\label{Lw}
\end{eqnarray}
where $G$ is the Newton gravitational constant, $\Lambda$ is a ({\it small}) cosmological term and 
$\kappa$ is an arbitrary constant of NLSUSY with the dimemsion (mass)${^{-2}}$.   
$w^a{}_\mu(x)$ $= e^a{}_\mu + t^a{}_\mu(\psi)$ and 
$w^{\mu}{_a}$ = $e^{\mu}{_a}
- t{^{\mu}}_a + t{^{\mu}}_{\rho} t{^{\rho}}_a - t{^{\mu}}_{\sigma}t{^{\sigma}}_{\rho} t{^{\rho}}_a  
+ t{^{\mu}}_{\kappa}t{^{\kappa}}_{\sigma}t{^{\sigma}}_{\rho}t{^{\rho}}_a  $ 
which terminates at ${\cal O}(t^{4})$ are the invertible unified vierbeins of new space-time.
{} $e^a{}_\mu$ is the ordinary vierbein of EGR for the local $SO(1,3)$ and    
$t^a{}_\mu(\psi)$ is the stress-energy-momentum tensor (i.e. the mimic vierbein) of 
NG fermion $\psi(x){}$ for the local $SL(2,C)$. 
(We call $\psi(x){}$  $superon$ as the hypothetical fundamental  spin $1/2$  particle 
constituting (carrying) the supercharge of the supercurrent \cite{KS4} of the global NLSUSY.) 
$\Omega(w)$ is the the unified scalar curvature of new space-time computed in terms of the unified vierbeins $w^a{}_\mu(x)$. 
Note that $e^a{}_\mu$ and $t^a{}_\mu(\psi)$ contribute equally to the curvature of space-time, 
which may be regarded as the Mach's principle in ultimate space-time. 
(The second index of mimic vierbein $t$, e.g. $\mu$ of $t^a{}_\mu$,  means the derivative $\partial_{\mu}$.)

$s_{\mu \nu} \equiv w^a{}_\mu \eta_{ab} w^b{}_\nu$ and 
$s^{\mu \nu}(x) \equiv w^\mu{}_a(x)w^{\nu a}(x)$ 
are unified metric tensors of new spacetime. 

NLSUSY GR action (\ref{SGM}) possesses promissing large symmetries 
isomorphic to $SO(N)$ ($SO(10)$) SP group \cite{ST3,ST6,ST4}; 
namely, $L_{\rm NLSUSYGR}(w)$ is invariant under 
\begin{eqnarray}
& & [{\rm new \ NLSUSY}] \otimes [{\rm local \ GL(4,R)}] 
\\
& & \otimes [{\rm local \ Lorentz}] \otimes \ [{\rm local \ spinor \ translation}] 
\end{eqnarray}
for spacetime symmetries 
and 
\begin{equation}
[{\rm global}\ SO(N)] \otimes [{\rm local}\ U(1)^N] 
\end{equation}
for internal symmetries in case of $N$ superons $\psi^i$, $(i = 1, 2, \cdots, N)$.  \\
For example,  $L_{\rm NLSUSYGR}(w)$ (\ref{SGM}) is invariant under the following NLSUSY transformations: 
\begin{equation}
\delta^{NL} \psi = {1 \over \kappa} \zeta 
- i \kappa \bar\zeta \gamma^\mu \psi \partial_\mu \psi,
\quad
\delta^{NL} e^a{}_\mu = i \kappa \bar\zeta \gamma^\rho \psi \partial_{[\mu} e^a{}_{\rho]},
\label{newsusy}
\end{equation} 
where $\zeta$ is a constant spinor and  $\partial_{[\mu} {e^{a}}_{\rho]} = 
\partial_{\mu}{e^{a}}_{\rho}-\partial_{\rho}{e^{a}}_{\mu}$. 
In fact, (\ref{newsusy}) induce the following GL(4R) transformations on the unified vierbein  $w{^a}_{\mu}$ 
\begin{equation}
\delta_{\zeta} {w^{a}}_{\mu} = \xi^{\nu} \partial_{\nu}{w^{a}}_{\mu} + \partial_{\mu} \xi^{\nu} {w^{a}}_{\nu}, 
\quad
\delta_{\zeta} s_{\mu\nu} = \xi^{\kappa} \partial_{\kappa}s_{\mu\nu} +  
\partial_{\mu} \xi^{\kappa} s_{\kappa\nu} 
+ \partial_{\nu} \xi^{\kappa} s_{\mu\kappa}, 
\label{newgl4r}
\end{equation} 
where  $\xi^{\mu} = - i \kappa \bar\zeta \gamma^\mu \psi$.  \\ 
$L_{\rm NLSUSYGR}(w)$ (\ref{SGM}) is  also invariant  under the following local Lorentz transformation: \\
on $w{^a}_{\mu}$ 
\begin{equation}
\delta_L w{^a}_{\mu}
= \epsilon{^a}_b w{^b}_{\mu}
\label{Lrw}
\end{equation}
or equivalently on $\psi$ and $e{^a}_{\mu}$ 
\begin{equation}
\delta_L \psi = - {i \over 2} \epsilon_{ab} 
      \sigma^{ab} \psi,     \quad
\delta_L {e^{a}}_{\mu} = \epsilon{^a}_b e{^b}_{\mu} 
      - {\kappa^2 \over 4} \varepsilon^{abcd} 
      \bar{\psi}\gamma_5 \gamma_d \psi 
      \partial_{\mu} \epsilon_{bc}, 
\label{newlorentz}
\end{equation}
with the local  parameter
$\epsilon_{ab} = (1/2) \epsilon_{[ab]}(x)$.   
The local Lorentz transformation forms a closed algebra, for example, on $e{^a}_{\mu}$ 
\begin{equation}
[\delta_{L_{1}}, \delta_{L_{2}}] e{^a}_{\mu} 
= \beta{^a}_b e{^b}_{\mu} 
- {\kappa^2 \over 4} \varepsilon^{abcd} \bar{\psi} 
\gamma_5 \gamma_d \psi 
\partial_{\mu} \beta_{bc}, 
\label{comLr1/2}
\end{equation}
where $\beta_{ab}=-\beta_{ba}$ is defined by 
$\beta_{ab} = \epsilon_{2ac}\epsilon{_1}{^c}_{b} -  \epsilon_{2bc}\epsilon{_1}{^c}_{a}$.  \par
The commutators of two new NLSUSY transformations (\ref{newsusy})  on $\psi$ and  ${e^{a}}_{\mu}$ 
are $GL(4R)$, i.e. new NLSUSY (\ref{newsusy}) is the square-root of $GL(4R)$; 
\begin{equation}
[\delta_{\zeta_1}, \delta_{\zeta_2}] \psi
= \Xi^{\mu} \partial_{\mu} \psi,
\quad
[\delta_{\zeta_1}, \delta_{\zeta_2}] e{^a}_{\mu}
= \Xi^{\rho} \partial_{\rho} e{^a}_{\mu}
+ e{^a}_{\rho} \partial_{\mu} \Xi^{\rho},
\label{com1/2-e}
\end{equation}
where 
$\Xi^{\mu} = 2 i \bar\zeta_1 \gamma^{\mu} \zeta_2 
      - \xi_1^{\rho} \xi_2^{\sigma} e{_a}^{\mu}
      \partial_{[\rho} e{^a}_{\sigma]}$. 
The algebra closes. 
%
The ordinary local $GL(4R)$ invariance is trivial by the construction.  \\  
Note that the no-go theorem is overcome (circumvented) in a sense that 
the nontivial $N$-extended SUSY gravity theory with $N > 8$ has been constructed in the NLSUSY invariant way. \par
New {\it empty} {} space-time for ${everything}$ described  
by NLSUSY GR $L_{\rm NLSUSYGR}(w)$ of the {\it vacuum} EH type 
is unstable due to NLSUSY structure of tangent space-time and 
decays (called {\it Big Decay} \cite{ST4}) spontaneously to 
ordinary Riemann space-time and superon (matter), which is described by 
the ordinary EH action with the cosmological constant $\Lambda$, 
NLSUSY action for $N$ superon (NG fermion) 
and their gravitational interactions. \par
The resulting action (called tentatively superon-graviton model (SGM) from the subsequent compositeness viewpoints) for $N=1$, 
which ignites Big Bang of the present observed universe,  is the following SGM action \cite{YC}; 
\begin{eqnarray}
& & L_{\rm SGM}(e, \psi) =  - {c^4\Lambda \over 16{\pi}G} e \vert w_{VA} \vert + {c^4 \over 16{\pi}G} e \vert w_{VA} \vert R(e) 
\nonumber \\[.5mm]
& & - {c^4 \over 16{\pi}G} e \vert w_{VA} \vert \left[ \ 2 t^{(\mu\nu)} R_{\mu\nu}(e) \right. 
\nonumber \\[.5mm]
& & + {1 \over 2} ( g^{\mu\nu}\partial^{\rho}\partial_{\rho}t_{(\mu\nu)}
- t_{(\mu\nu)}\partial^{\rho}\partial_{\rho}g^{\mu\nu}       
\nonumber \\[.5mm]
& & + g^{\mu\nu}\partial^{\rho}t_{(\mu\sigma)}\partial^{\sigma}g_{\rho\nu}
- 2g^{\mu\nu}\partial^{\rho}t_{(\mu\nu)}\partial^{\sigma}g_{\rho\sigma}
- g^{\mu\nu}g^{\rho\sigma}\partial^{\kappa}t_{(\rho\sigma)}\partial_{\kappa}g_{\mu\nu} )     
\nonumber \\[.5mm]
& & + ({t^{\mu}}_{\rho}t^{\rho\nu}+{t^{\nu}}_{\rho}t^{\rho\mu}+t^{\mu\rho}{t^{\nu}}_{\rho})R_{\mu \nu}(e)    
\nonumber \\[.5mm]
& & - 2 t^{(\mu\rho)}{t^{(\nu}}_{\rho)}R_{\mu\nu} - t^{(\mu \rho)} t^{(\nu \sigma)} R_{\mu \nu \rho \sigma}(e)  
\nonumber \\[.5mm]
& & - {1 \over 2}t^{(\mu\nu)}( g^{\rho\sigma}\partial_{\mu}\partial_{\nu}t_{(\rho\sigma)} 
- g^{\rho \sigma} \partial_{\rho}\partial_{\mu}t_{(\sigma\nu)} + \cdots ) 
\nonumber \\[.5mm]
& & \left. + {\cal O}(t^3) + \cdots  \ \right],
\label{L-exp}
\end{eqnarray}
where $(\psi)^{n} \equiv 0$ ($n \geq 5$ for $N = 1$), $e = \det{e^{a}}_{\mu}$, $t^{(\mu\nu)}=t^{\mu\nu}+t^{\nu\mu}$, 
$t_{(\mu\nu)}=t_{\mu\nu}+t_{\nu\mu}$, and $\vert w_{VA} \vert = \det{w^{a}}_{b} = \det(\delta^{a}{_b}+ t^{a}{_b})$ 
is the flat space NLSUSY action of VA \cite{VA} 
containing up to ${\cal O}(t^4)$ and $R(e)$, $R_{\mu\nu}(e)$ and $R_{\mu\nu\rho\sigma}(e)$ 
are the ordinary scalar curvature, the Ricci tensor and the Riemann curvature tensor of GR. 
Remarkably we can observe that  the first term reduces to NLSUSY action \cite{VA}, 
i.e. the arbitrary constant ${\kappa}$ of NLSUSY is now fixed to  
\begin{equation}
\kappa^{-2} = {c^4 \over {8 \pi G}} \Lambda
\label{kappa}
\end{equation}
in the Riemann-flat $e{_a}^{\mu}(x) \rightarrow \delta{_a}^{\mu}$ space-time 
and the second term contains the familiar EH action of GR. 
These expansions describe the complementary relation of graviton and superons (matter), 
i.e. Mach's principle is encoded. 
Note that  NLSUSY GR (\ref{SGM}) and SGM (\ref{L-exp}) possess different asymptotic flat space-time, i.e. 
SGM-flat $w{_a}^{\mu}(x) \rightarrow \delta{_a}^{\mu}$ space-time and 
Riemann-flat  $e{_a}^{\mu}(x) \rightarrow \delta{_a}^{\mu}$ space-time, respectively. \par
$L_{\rm SGM}(e, \psi)$ (\ref{L-exp}) can be rewritten as the following famlliar form  
\begin{equation}
L_{\rm SGM}(e,\psi) = {c^{4} \over 16{\pi}G}\vert e \vert \{ R(e) - \Lambda + \tilde T(e, \psi) \},
\label{SGMR}
\end{equation}
%
where $R(e)$ is the scalar curvature of ordinary EH action in Riemann space-time and 
$\tilde T(e,\psi)$ represents the kinetic term and the gravitational interaction of superons. \par
We think that the geometric arguments of EGR principle 
has been generalized naturally, which accomodates {\it geometrically} spin $1/2$ matter 
as NG fermion accompanying spontaneous SUSY breaking encoded on tangent space-time as NLSUSY.  
Therefore the black hole as a singularity of space-time in EGR is an interesting object 
to be studied in the picture of  NLSUSY GR.    \par
We have shown qualitatively that NLSUSY GR may potentially describe a new paradigm (SGM) 
for the SUSY unification of space-time and matter,  
where particular SUSY composites composed of superons for all (observed) particles except the graviton emerges 
as an ultimate feature of nature behind the familiar LSUSY models (MSSM, SUSY GUTs) \cite{KS3,STS} and SM as well. 
That is, all (observed) low energy particles may be eigenstates of $SO(N)$ SP expressed uniquely 
as the SUSY composites (eigen states) of $N$ superons. We examine these possibilities in the next section.
\section{NL/L SUSY Relation and Physical Implications}
Due to the high nonlinearity of the SGM action we have not yet succeeded in extracting directly 
(low energy) physical meanings of SGM on curved Riemann space-time.     %

However, considering that the SGM action reduces essentially to the $N$-extended NLSUSY action 
with ${\kappa^{2} = ({c^{4}\Lambda \over 8{\pi}G}})^{-1}$  
in asymptotic Riemann-flat $(e^a{}_\mu \rightarrow \delta^a_\mu)$ space-time, 
it is interesting from the viewpoint of the low energy physics on the local coordinate system 
to  find the $N$-extended LSUSY theory equivalent (related) to the $N$-extended NLSUSY model. 
Through the linearization of the  $N$-extended NLSUSY model, the relation between $N $ LSUSY ${\it free}$ theory for LSUSY supermultiplet 
and $N$ NLSUSY VA  model for $N$ NG fermion is demonstrated by many authors for $N = 1$ \cite{IK,R,UZ,STT1} 
and for $N = 2$ case \cite{STT2} by heuristic arguments and by the systematic superfield arguments as well. 

We have shown explicitly by the heuristic arguments for simplicity 
in two space-time dimensions ($d = 2$) \cite{ST2,ST5} 
that $N = 2$ LSUSY interacting QED is equivalent (related) to $N = 2$ NLSUSY model  
in a sense that analogous SUSY invariant relations hold, 
i.e. each field of LSUSY supermultiplet are expressed uniquely in terms of NLSUSY NG fermions 
by the arguments on SUSY transformation.
Note that the minimal realistic SUSY QED in SGM composite scenario is given by $N = 2$ SUSY \cite{STT2}.

In establishing {\it NL/L SUSY relation} each field of LSUSY supermultiplet 
is expressed uniqely as the composite of NG fermions of NLSUSY, 
which are called  {\it SUSY invariant relations}. 
Consequently we are tempted to imagine some composite structure (far) behind 
the familiar LSUSY unified models, e.g. MSSM and SUSY GUT.  
In this paper we study explicitly the vacuum structure of $N = 2$ LSUSY QED in the SGM scenario in $d = 2$ \cite{ST5}. 

$N = 2$ NLSUSY action for two superons (NG fermions) $\psi^i\ (i, j, \cdots = 1, 2)$ in $d = 2$ is written 
as follows, 
\begin{eqnarray}
& & L_{N=2{\rm NLSUSY}}
\nonumber \\[.5mm]
& & = -{1 \over {2 \kappa^2}} \vert w \vert
\nonumber \\[.5mm]
& & = - {1 \over {2 \kappa^2}} 
\left\{ 1 + t^a{}_a + {1 \over 2!}(t^a{}_a t^b{}_b - t^a{}_b t^b{}_a) 
\right\} 
\nonumber \\[.5mm]
& & = - {1 \over {2 \kappa^2}} 
\left\{ 1 - i \kappa^2 \bar\psi^i \!\!\not\!\partial \psi^i \right. 
\nonumber \\[.5mm]
& & 
\left. - {1 \over 2} \kappa^4 
( \bar\psi^i \!\!\not\!\partial \psi^i \bar\psi^j \!\!\not\!\partial \psi^j 
- \bar\psi^i \gamma^a \partial_b \psi^i \bar\psi^j \gamma^b \partial_a \psi^j ) 
\right\} 
\label{VAaction2}
\end{eqnarray}
where $\kappa$ is a constant whose dimension is $({\rm mass})^{-1}$ and 
%
$\vert w \vert = \det(w^a{}_b) = \det(\delta^a_b + t^a{}_b)$, 
$t^a{}_b = - i \kappa^2 \bar\psi^i \gamma^a \partial_b \psi^i$, 
which is invariant under $N = 2$ NLSUSY transformation, 
\begin{equation}
\delta_\zeta \psi^j ={ {1 \over \kappa} \zeta^j }
- i \kappa \bar\zeta^k \gamma^a \psi^k \partial_a \psi^j. 
\label{nlsusytr}
\end{equation}
While, the helicity states contained formally in ($d = 2$) $N = 2$ LSUSY QED are 
the vector supermultiplet containing $U(1)$ gauge field 
%
\[\left(\begin{array}{c}
      +{1} \\
\begin{array}{cc}
 +{1 \over 2}, \  +{1 \over 2} 
\end{array} \\
0
\end{array}  \right) + [{\rm CPT\ conjugate}], \]
and the scalar supermultiplet for matter fields  
%
\[\left(\begin{array}{c}
      +{1 \over 2} \\
\begin{array}{cc}
 0 ,\  0 
\end{array} \\
-{1 \over 2}
\end{array}  \right) + [{\rm CPT\ conjugate}]. \]
The most general $N = 2$ LSUSY QED action for the massless case in $d = 2$, 
is written as follows \cite{ST5}, 
\begin{eqnarray}
& & L_{N=2{\rm SUSYQED}} 
\nonumber \\[.5mm]
& & = - {1 \over 4} (F_{ab})^2 
+ {i \over 2} \bar\lambda^i \!\!\not\!\partial \lambda^i 
+ {1 \over 2} (\partial_a A)^2 
+ {1 \over 2} (\partial_a \phi)^2 
+ {1 \over 2} D^2 
\nonumber \\[.5mm]
& & 
- {1 \over \kappa} \xi D 
+ {i \over 2} \bar\chi \!\!\not\!\partial \chi 
+ {1 \over 2} (\partial_a B^i)^2 
+ {i \over 2} \bar\nu \!\!\not\!\partial \nu 
+ {1 \over 2} (F^i)^2 
\nonumber \\[.5mm]
& & 
+ f ( A \bar\lambda^i \lambda^i + \epsilon^{ij} \phi \bar\lambda^i \gamma_5 \lambda^j 
- A^2 D + \phi^2 D + \epsilon^{ab} A \phi F_{ab} ) 
\nonumber \\[.5mm]
& & 
+ e \left\{ i v_a \bar\chi \gamma^a \nu 
- \epsilon^{ij} v^a B^i \partial_a B^j 
+ \bar\lambda^i \chi B^i 
+ \epsilon^{ij} \bar\lambda^i \nu B^j 
\right. 
\nonumber \\[.5mm]
& & 
\left. 
- {1 \over 2} D (B^i)^2 
+ {1 \over 2} (\bar\chi \chi + \bar\nu \nu) A 
- \bar\chi \gamma_5 \nu \phi \right\}
\nonumber \\[.5mm]
& & 
+ {1 \over 2} e^2 (v_a{}^2 - A^2 - \phi^2) (B^i)^2, 
\label{L2action}
\end{eqnarray}
where  $(v^a, \lambda^i, A, \phi, D)$ ($F_{ab} = \partial_a v_b - \partial_b v_a$) 
is the off-shell vector supermultiplet containing $v^a$ for a $U(1)$ vector field, 
$\lambda^i$ for doublet (Majorana) fermions 
$A$ for a scalar field in addition to $\phi$ for another scalar field 
and $D$ for an auxiliary scalar field, 
while ($\chi$, $B^i$, $\nu$, $F^i$) is off-shell scalar supermultiplet containing 
$(\chi, \nu)$ for two (Majorana) fermions, 
$B^i$ for doublet scalar fields and $F^i$ for auxiliary scalar fields. 
The linear term of $F$ is forbidden by the gauge invariance \cite{ST5}. 
Also $\xi$ is an arbitrary demensionless parameter giving a magnitude of SUSY breaking mass, 
and $f$ and $e$ are Yukawa and gauge coupling constants with the dimension (mass)$^1$, respectively. \par
$N = 2$ LSUSY QED action (\ref{L2action}) is invariant under the following LSUSY transformations  
parametrized by $\zeta^i$, 
\begin{eqnarray}
& &
\delta_\zeta v^a = - i \epsilon^{ij} \bar\zeta^i \gamma^a \lambda^j, 
\nonumber \\[.5mm]
& &
\delta_\zeta \lambda^i 
= (D - i \!\!\not\!\partial A) \zeta^i 
+ {1 \over 2} \epsilon^{ab} \epsilon^{ij} F_{ab} \gamma_5 \zeta^j 
- i \epsilon^{ij} \gamma_5 \!\!\not\!\partial \phi \zeta^j, 
\nonumber \\[.5mm]
& &
\delta_\zeta A = \bar\zeta^i \lambda^i, 
\nonumber \\[.5mm]
& &
\delta_\zeta \phi = - \epsilon^{ij} \bar\zeta^i \gamma_5 \lambda^j, 
\nonumber \\[.5mm]
& &
\delta_\zeta D = - i \bar\zeta^i \!\!\not\!\partial \lambda^i, 
\label{VLSUSY}
\end{eqnarray}
for the vector multiplet and 
\begin{eqnarray}
\delta_\zeta \chi 
& = & (F^i - i \!\!\not\!\partial B^i) \zeta^i - e \epsilon^{ij} V^i B^j, 
\nonumber \\[.5mm]
\delta_\zeta B^i 
& = & \bar\zeta^i \chi - \epsilon^{ij} \bar\zeta^j \nu, 
\nonumber \\[.5mm]
\delta_\zeta \nu 
& = & \epsilon^{ij} (F^i + i \!\!\not\!\partial B^i) \zeta^j + e V^i B^i, 
\nonumber \\[.5mm]
\delta_\zeta F^i 
& = & - i \bar\zeta^i \!\!\not\!\partial \chi 
- i \epsilon^{ij} \bar\zeta^j \!\!\not\!\partial \nu 
\nonumber \\[.5mm]
& &
- e \{ \epsilon^{ij} \bar V^j \chi - \bar V^i \nu 
+ (\bar\zeta^i \lambda^j + \bar\zeta^j \lambda^i) B^j 
- \bar\zeta^j \lambda^j B^i \} 
\label{SLSUSYg}
\end{eqnarray}
%
with $V^i = i v_a \gamma^a \zeta^i - \epsilon^{ij} A \zeta^j - \phi \gamma_5 \zeta^i$ for the scalar multiplet.  \par 
$N = 2$ LSUSY QED action (\ref{L2action}) and (\ref{SLSUSYg}) can be rewritten as the familiar manifestly covariant form 
in terms of the complex quantities defined by 
\begin{equation}
\chi_D = {1 \over \sqrt{2}} (\chi + i \nu), 
\ \ \ B = {1 \over \sqrt{2}} (B^1 + i B^2), 
\ \ \ F = {1 \over \sqrt{2}} (F^1 - i F^2).  
\label{cfields}
\end{equation}
The resulting action is manifestly invariant under the  local $U(1)$: 
\begin{eqnarray}
& & 
(\chi_D, B, F) \ \ \rightarrow \ \ (\chi'_D, B', F')(x) = e^{i \Omega(x)} (\chi_D, B, F)(x), 
\nonumber \\[.5mm]
& & 
v_a \ \ \rightarrow \ \ v'_a(x)  = v_a(x) + {1 \over e} \partial_a \Omega(x). 
\label{u1}
\end{eqnarray}
(For further details see ref.\cite{ST5}.)     \par 
For extracting the low energy particle physics contents of $N = 2$ SGM (NLSUSY GR) 
we adopt Riemann-flat asymptotic space-time, where $N = 2$ SGM reduces to 
essentially $N = 2$ NLSUSY action which is related (equivalent) to $N = 2$ SUSY QED action, 
which we call {\it NL/L SUSY relation}, i.e. for $e^a{}_\mu \rightarrow \delta^a{}_\mu$
%
\begin{eqnarray}
& & 
L_{N=2{\rm SGM}} 
{\longrightarrow} 
L_{N=2{\rm NLSUSY}} + [{\rm suface\ terms}]
= L_{N=2{\rm SUSYQED}}. 
\label{equality}
\end{eqnarray}
%
The NL/L SUSY relation of the two theories are shown explicitly by substituting 
the following generalized SUSY invariant relations \cite{ST5} into the LSUSY QED theory. \\
The SUSY invariant relations for the vector supermultiplet $(v^a, \lambda^i, A, \phi, D)$ are 
\begin{eqnarray}
& & 
v^a = - {i \over 2} \xi \kappa \epsilon^{ij} 
\bar\psi^i \gamma^a \psi^j \vert w \vert, 
\nonumber \\[.5mm]
& & 
\lambda^i = \xi \left[ \psi^i \vert w \vert 
- {i \over 2} \kappa^2 \partial_a 
\{ \gamma^a \psi^i \bar\psi^j \psi^j \vert w \vert
 \} \right], 
\nonumber \\[.5mm]
& & 
A = {1 \over 2} \xi \kappa \bar\psi^i \psi^i \vert w \vert, 
\nonumber \\[.5mm]
& & 
\phi = - {1 \over 2} \xi \kappa \epsilon^{ij} \bar\psi^i \gamma_5 \psi^j 
\vert w \vert, 
\nonumber \\[.5mm]
& & 
D = {\xi \over \kappa} \vert w \vert 
- {1 \over 8} \xi \kappa^3 
\partial_a \partial^a ( \bar\psi^i \psi^i \bar\psi^j \psi^j \vert w \vert),  
\label{SSUSYinv1}
\end{eqnarray}
while for the scalar supermultiplet $(\chi, B^i, \nu, F^i)$ the relaxed SUSY invariant relations are   
\begin{eqnarray}
& & 
\chi = \xi^i \left[ \psi^i \vert w \vert 
+ {i \over 2} \kappa^2 \partial_a 
\{ \gamma^a \psi^i \bar\psi^j \psi^j \vert w \vert
 \} \right], 
\nonumber \\[.5mm]
& & 
B^i = - \kappa \left( {1 \over 2} \xi^i \bar\psi^j \psi^j 
- \xi^j \bar\psi^i \psi^j \right) \vert w \vert, 
\nonumber \\[.5mm]
& & 
\nu = \xi^i \epsilon^{ij} \left[ \psi^j \vert w \vert 
+ {i \over 2} \kappa^2 \partial_a 
\{ \gamma^a \psi^j \bar\psi^k \psi^k \vert w \vert
\} \right], 
\nonumber \\[.5mm]
& & 
F^i = {1 \over \kappa} \xi^i \left\{ \vert w \vert 
+ {1 \over 8} \kappa^3 
\partial_a \partial^a ( \bar\psi^j \psi^j \bar\psi^k \psi^k \vert w \vert )
\right\} 
\nonumber \\[.5mm]
& & 
- i \kappa \xi^j \partial_a ( \bar\psi^i \gamma^a \psi^j \vert w \vert ) 
- {1 \over 4} e \kappa^2 \xi \xi^i \bar\psi^j \psi^j \bar\psi^k \psi^k \vert w \vert. 
\label{SSUSYinv}
\end{eqnarray}
$N=2$ SUSYQED action (\ref{L2action}) is invariant under the variations of the 
LSUSY component fields of two supermultiplets, which are  induced by the NLSUSY transformations 
of the superon $\psi^i$ contained in the modified SUSY invariant relations (\ref{SSUSYinv1}) and 
(\ref{SSUSYinv}).
Furthermore substituting these SUSY invariant relations into the $N=2$ SUSYQED action (\ref{L2action})
we can show directly that $N=2$ SUSYQED action (\ref{L2action}) is related (reduced) to $N = 2$ NLSUSY action 
provided $\xi^{2}-(\xi^{i})^{2}=1$.  \\
As for the LSUSY transformations  (\ref{VLSUSY}) and (\ref{SLSUSYg}), 
the SUSY invariant relations (\ref{SSUSYinv1}) reproduce 
the familiar LSUSY transformations (\ref{VLSUSY}) under the NLSUSY transformations 
on the contained superons $\psi^i$. 
The equality in (\ref{equality}) and the SUSY invariant relations (\ref{SSUSYinv1}) and (\ref{SSUSYinv}) 
are called NL/L SUSY relation.   \par
While the modified SUSY invariant relations (\ref{SSUSYinv}) for the scalar supermultiplet mean that 
the familiar LSUSY transformations (\ref{SLSUSYg}) on the above scalar supermultiplet 
are not reproduced by the NLSUSY transformations of the contained superons $\psi^i$ 
but modified by the (contact four-fermion) gauge interaction terms in the above heuristic arguments. 
It is interesting that the four-fermion self-interaction term 
appearing only in the auxiliary fields $F^i$ 
is the origin of the familiar local $U(1)$ gauge symmetry of LSUSY theory, 
which makes apparently the SUSY invariant relations modified. 
The redefinitions and the generalizations  of the auxiliary fields in the supermultiplet 
solve these unpleasant situations \cite{ST7}, 
i.e. by the systematic arguments of NL/L SUSY relations in terms of the superfields 
the familiar LSUSY transformations on the generalized scalar supermultiplet are reproduced and 
NL/L SUSY relation holds. 
Note that in the linearized theory the commutator algebra does not contain U(1) gauge transformation 
even for the vector U(1) gauge field \cite{STT2}.   \par
Now we study the vacuum structure of $N = 2$ SUSY QED action (\ref{L2action}) \cite{STL}. 
The vacuum is determined by the minimum of the potential $V(A, \phi, B^i, D)$, 
\footnote{
The terms, ${1 \over 2} e^2 (A^2 + \phi^2) (B^i)^2$, 
should be added to Eqs.(\ref{pot1}) and (\ref{pot2}) 
but they do not change the final results. 
}
\begin{equation}
V(A, \phi, B^i, D) 
=  - {1 \over 2} D^2 + \left\{ {\xi \over \kappa} 
+ f(A^2 - \phi^2) + {1 \over 2} e (B^i)^2 \right\} D.  
\label{pot1}
\end{equation}
Substituting the solution of the equation of motion for the auxiliary field $D$
we obtain 
\begin{equation}
V(A, \phi, B^i) = {1 \over 2} f^2 \left\{ A^2 - \phi^2 + {e \over 2f} (B^i)^2 
+ {\xi \over {f \kappa}} \right\}^2 \ge 0. 
\label{pot2}
\end{equation}
The configurations of the fields corresponding to the vacua in $(A, \phi, B^i)$-space, 
which are $SO(1,3)$ or $SO(3,1)$ invariant,   
are classified according to the signatures of the parameters $e, f, \xi, \kappa$ 
as follows: \\
(I) For $ef < 0$, \ \ ${\xi \over {f \kappa}} < 0$ case, 
%
\begin{equation}
A^2 - \phi^2 - (\tilde B^i)^2 = k^2. 
\ \ \ \left( \tilde B^i = \sqrt{-e \over 2f} B^i, \ \ 
k^2 = {-\xi \over {f \kappa}} \right) 
\end{equation}
%
(II) For $ef > 0$, \ \ ${\xi \over {f \kappa}} < 0$ case, 
%
\begin{equation}
A^2 - \phi^2 + (\tilde B^i)^2 = k^2. 
\ \ \ \left( \tilde B^i = \sqrt{{e \over 2f}} B^i, \ \ 
k^2 = {-\xi \over {f \kappa}} \right) 
\end{equation}
%
(III) For $ef < 0$, \ \ ${\xi \over {f \kappa}} > 0$ case, 
%
\begin{equation}
- A^2 + \phi^2 + (\tilde B^i)^2 = k^2. 
\ \ \ \left( \tilde B^i = \sqrt{-e \over 2f} B^i, \ \ 
k^2 =  {\xi \over {f \kappa}} \right) 
\end{equation}
%
(IV) For $ef > 0$, \ \ ${\xi \over {f \kappa}}> 0$ case, 
%
\begin{equation}
- A^2 + \phi^2 - (\tilde B^i)^2 = k^2. 
\ \ \ \left( \tilde B^i = \sqrt{{e \over 2f}} B^i, \ \ 
k^2 =  {\xi \over {f \kappa}} \right) 
\end{equation}
%
We find that the vacua (I) and (IV) with $SO(1,3)$ isometry in $(A, \phi, B^i)$-space are unphysical, 
for they produce the pathological wrong sign kinetic terms for the fields induced around the vacuum. 

As for the cases (II) and (III) we perform the similar arguments as shown below 
and find that two different physical vacua appear. 
The physical particle spectrum is obtained by expanding the field $(A, \phi, B^i)$ around the vacuum 
with $SO(3,1)$ isometry.              \par
For case (II), the following expressions (IIa) and (IIb) are considered: \\
Case (IIa) 
\[
\begin{array}{lll}
A & = (k + \rho)\sin\theta \cosh\omega & {}      
\\
\phi & = (k + \rho) \sinh\omega & {}
\\
\tilde B^1 & = (k + \rho) \cos\theta \cos\varphi \cosh\omega & {}
\\
\tilde B^2 & = (k + \rho) \cos\theta \sin\varphi \cosh\omega. & {}
\end{array}
\]
and \\
Case (IIb) 
\[
\begin{array}{lll}
A &\!\!\! = - (k + \rho) \cos\theta \cos\varphi \cosh\omega &\!\!\! {}
\\
\phi &\!\!\! = (k + \rho) \sinh\omega &\!\!\! {}
\\
\tilde B^1 &\!\!\! = (k + \rho) \sin\theta \cosh\omega &\!\!\ {}
\\
\tilde B^2 &\!\!\! = (k + \rho) \cos\theta \sin\varphi \cosh\omega &\!\!\! {}
\end{array}
\]
Note that for the case (III) the arguments are the same by exchanging $A$ and $\phi$, 
which we call (IIIa) and (IIIb). 
Substituting these expressions into $ L_{N=2{\rm SUSYQED}}$ $(A, \phi, B^i)$ and 
expanding the action around the vacuum configuration  we obtain the physical particle contents. 
For the cases (IIa) and (IIIa) we obtain 
\begin{eqnarray}
& & 
L_{N=2{\rm SUSYQED}} 
\nonumber \\[.5mm]
& & = {1 \over 2} \{ (\partial_a \rho)^2 - 2 (ef) k^2 \rho^2 \} 
\nonumber \\[.5mm]
& & 
+ {1 \over 2} \{ (\partial_a \theta)^2 + (\partial_a \omega)^2 - 2 (ef) k^2 (\theta^2 + \omega^2) \} 
\nonumber \\[.5mm]
& & 
+ {1 \over 2} (\partial_a \varphi)^2 
\nonumber \\[.5mm]
& & 
- {1 \over 4} (F_{ab})^2 + (ef) k^2 v_a^2 
\nonumber \\[.5mm]
& & 
+ {i \over 2} \bar\lambda^i \!\!\not\!\partial \lambda^i 
+ {i \over 2} \bar\chi \!\!\not\!\partial \chi 
+ {i \over 2} \bar\nu \!\!\not\!\partial \nu 
\nonumber \\[.5mm]
& & 
+ \sqrt{2ef} (\bar\lambda^1 \chi - \bar\lambda^2 \nu) 
+ \cdots, 
\end{eqnarray}
and the consequent mass genaration 
\begin{eqnarray}
& & 
m_\rho^2 = m_\theta^2 = m_\omega^2 = m_{v_a}^2 = 2 (ef) k^2  
= - {{2 \xi e} \over \kappa}, 
\nonumber \\[.5mm]
& & 
m_{\lambda^{i}} = m_{\chi} =  m_{\nu}= m_{\varphi}= 0. 
\nonumber \\[.5mm]
& & 
\end{eqnarray}
Note that $\varphi$ is NG boson for the spontaneous breaking of $U(1)$ symmetry, i.e. the $U(1)$ phase of $B$,  
and totally gauged away by the Higgs-Kibble mechanism with $\Omega(x) = \sqrt{ e \kappa/2}\varphi(x)$ 
for $U(1)$ gauge (\ref{u1}). 
The vacuum breaks both SUSY and the local $U(1)$ spontaneously. 
All bosons have the same mass and  remarkably all fermions remain massless.
$\lambda^{i}$ transforming inhomogeneouly 
$\delta \lambda^{i}={\xi \over \kappa}\zeta^{i}+ \cdots$ in the true vacuum 
are  NG fermions for the spontaneous $N=2$ SUSY breaking. 
The physical implication of the off-diagonal mass terms 
$\sqrt{2ef} (\bar\lambda^1 \chi - \bar\lambda^2 \nu) 
= \sqrt{2ef} (\bar\chi_{\rm D} \lambda + \bar\lambda \chi_{\rm D})$ for fermions is 
unclear (pathologocal?) and deserves further investigations, which would induce the mixings of fermions 
and/or the lepton (baryon) flavour  violations. \par  
By the similar computations for (IIb) and (IIIb) we obtain 
\begin{eqnarray}
& & 
L_{N=2{\rm SUSYQED}} 
\nonumber \\[.5mm]
& & = {1 \over 2} \{ (\partial_a \rho)^2 - 4 f^2 k^2 \rho^2 \} 
\nonumber \\[.5mm]
& & 
+ {1 \over 2} \{ (\partial_a \theta)^2 + (\partial_a \varphi)^2 
- e^2 k^2 (\theta^2 + \varphi^2) \} 
\nonumber \\[.5mm]
& & 
+ {1 \over 2} (\partial_a \omega)^2 
\nonumber \\[.5mm]
& & 
- {1 \over 4} (F_{ab})^2 
\nonumber \\[.5mm]
& & 
+ {1 \over 2} (i \bar\lambda^i \!\!\not\!\partial \lambda^i 
- 2 f k \bar\lambda^i \lambda^i) 
\nonumber \\[.5mm]
& & 
+ {1 \over 2} \{ i (\bar\chi \!\!\not\!\partial \chi + \bar\nu \!\!\not\!\partial \nu) 
- e k (\bar\chi \chi + \bar\nu \nu) \} 
+ \cdots. 
\label{VACb}
\end{eqnarray}
and the following mass spectrum which indicates the spontaneous breakdown of $N=2$ SUSY; 
\begin{eqnarray}
& & 
m_\rho^2 = m_{\lambda^i}^2 = 4 f^2 k^2 = {-{4 \xi f} \over \kappa}, 
\nonumber \\[.5mm]
& & 
m_\theta^2 = m_\varphi^2 = m_\chi^2 = m_\nu^2 
= e^2 k^2 = {{-\xi e^2} \over {\kappa f}}, 
\nonumber \\[.5mm]
& & 
m_{v_{a}} = m_{\omega} = 0,
\end{eqnarray}
which can produce {mass hierarchy} by the factor 
%
${e \over f}$.  
Interestingly all fermions acquire masses through the spontaneous SUSY breaking. 
The local $U(1)$ gauge symmetry is not broken. 
The massless scalar $\omega$ is a NG boson for the degeneracy of the vacuum in $(A,\tilde B_{2})$-space, 
which is gauged away provided the local gauge symmetry between the vector and the scalar multiplet is introduced. \par
From these arguments we conclude that $N = 2$ SUSY QED is equivalent (related) to $N = 2$ NLSUSY action, 
which is the matter sector of $N = 2$ SGM produced by Big Decay (phase transition) of $N = 2$ NLSUSY GR (new space-time). 
It possesses two different vacua, the type (a) and (b) in the $SO(3,1)$ isometry of (II) and (III). \par 
The resulting models describe; \\
for the type (a); two charged chiral fermions ($\psi_L{}^{cj} \sim (\tilde \chi_{DL}, \tilde \nu_{DL})$) $(j = 1, 2)$, 
two neutral chiral fermions ($\lambda_L{}^j \sim \tilde \lambda_{DL}^j$), 
one massive vector ($v_a$), 
one charged massive scalar ($\phi^c \sim \theta + i \omega$),  
and one massive scalar ($\phi^0 \sim \rho$), 
where ($\chi, \nu, \lambda^{i}$) are written by left-handed Dirac fields  and  \\ 
for the type (b); 
one charged Dirac fermion ($\psi_D{}^c \sim \chi + i \nu$), 
one neutral (Dirac) fermion ($ \lambda_D{}^0 \sim \lambda^1 - i \lambda^2$), 
one massless vector (a photon) ($v_a$), 
one charged scalar ($\phi^c \sim \theta + i \varphi$) 
and one neutral complex scalar ($\phi^{0c} \sim \rho + i \omega$), \\
which are the composites of superons. 
\section{Cosmology of NLSUSY GR}
Now we discuss the cosmological implications of $N=2$ NLSUSY GR (or $N=2$ SGM from the composite viewpoints). 
NLSUSY GR space-time of (\ref{SGM}) is unstable and spontaneously breaks down to Riemann space-time 
with superon (massless NG fermion) matter of (\ref{SGMR}), which may be the birth of the present universe 
(space-time and matter) by the quantum effect Big Decay in advance of the  inflation and/or the Big Bang. 
The variation of (\ref{SGMR}) with respect to  ${e^{a}}_{\mu}$ gives 
the equation of motion for ${e^{a}}_{\mu}$ recasted as follows; 
\begin{equation}
R_{\mu\nu}(e)-{1 \over 2}g_{\mu\nu}R(e) = 
{8{\pi}G \over c^{4}} \left\{ \tilde T_{\mu\nu}(e,{\psi}) 
- g_{\mu\nu}{c^{4}\Lambda \over 8{\pi}G} \right\}, 
\label{SGMEQ}
\end{equation}
where $\tilde T_{\mu\nu}(e,{\psi})$ abbreviates the stress-energy-momentum of superon (NG fermion) matter 
including the gravitational interactions. 
Note that $-{c^{4}\Lambda \over 8{\pi}G}$ can be interpreted as 
{\it the negative energy density of empty spacetime}, 
i.e. {\it the dark energy density ${{\rho}_{D}}$}. 
(The negative sign determined uniquely and produces the correct kinetic term of superons in NLSUSY.) 

While, on tangent spacetime, the low energy theorem of the particle physics 
gives the following superon (massless NG fermion)-vacuum coupling 
\begin{equation}
<{\psi^j}_{\alpha}(q) \vert {J^{k\mu}}_{\beta} \vert 0> = 
i\sqrt{c^{4}\Lambda \over 8{\pi}G}(\gamma^{\mu})_{\alpha\beta} \delta^{jk} e^{iqx}+ \cdots,  
\label{LETH}
\end{equation}
where ${{J^{k\mu}}= i\sqrt{c^{4}\Lambda \over 8{\pi}G} \gamma^{\mu}\psi^k + \cdots}$ $(j, k = 1, 2)$ 
is the conserved supercurrent obtained by applying the Noether theorem \cite{KS4}         
and $\sqrt{c^{4}\Lambda \over 8{\pi}G}$ is {\it the coupling constant $g_{sv}$ 
of superon with the vacuum via the supercurrent}. 
Further we have seen in the preceding section that the right hand side of (\ref{SGMEQ}) for $N = 2$  
is essentially $N = 2$ NLSUSY VA action. 
And it is equivalent to the broken $N=2$ LSUSYQED action (\ref{L2action}) 
with the vacuum expectation value of the auxiliary field (Fayet-Iliopoulos term). 
For the case (b) it gives the SUSY breaking masse, 
\begin{equation}
{M_{SUSY}}^{2} \sim <D> \sim \sqrt{c^{4}\Lambda \over 8{\pi}G}, 
\label{MASS1}
\end{equation}
to the component fields of the (massless) LSUSY supermultiplet, provided $-f\xi \sim {\cal O}(1)$. 
We find  NLSUSY GR (SGM) scenario gives interesting relations among the important quantities 
of the cosmology and the low energy particle physics, i.e. 
\begin{equation}
\rho_{D} \sim {c^{4}\Lambda \over 8{\pi}G} \sim <D>^{2} \sim  g_{sv}{^2}, 
\label{MASS2}
\end{equation}
Suppose that in the (low energy) LSUSY supermultiplet 
the stable and the lightest massive particle retains the mass of the order of the spontaneous SUSY breaking.  \\
And if we identify the neutrino with such a particle and with $\lambda^{i}(x)$, 
i.e.
\begin{equation}
{{m_{\nu}}^{2} \sim \sqrt{c^{4}\Lambda \over 8{\pi}G}}, 
\label{numass}
\end{equation}
then SGM predicts remarkably the observed value of the (dark) energy density of the universe and 
explains naturally  the observed mysterious numerical relations between $m_{\nu}$ and $\rho^{obs}_{D}$:
\begin{equation}
{\rho^{obs}_{D}} \sim (10^{-12}GeV)^{4} \sim {m_{\nu}}^{4}{} \ (\sim  g_{sv}{}^2).
\label{DENSITY}
\end{equation}
The tiny neutrino mass is the direct evidence of SUSY (breaking), i.e. 
the spontaneous phase transition of SGM spacetime. 
The large mass scales and the non-abelian gauge symmetry necessary for building the realistic 
and interacting broken LSUSY model will appear by the extension to the large $N$ SUSY and 
by the linearization of $\tilde T_{\mu\nu}(e,{\psi})$ 
which contains the mass scale $\Lambda^{-1}$ in the higher order with $\psi$ \cite{ST6}.

\section{Conclusions}
We have proposed a new paradigm for describing the unity of nature, 
where the ultimate shape of nature is  new uastable (empty) space-time 
described by $L_{\rm NLSUSYGR}(w)$. $L_{\rm NLSUSYGR}(w)$ decays spontaneously to 
ordinary Riemann space-time with massless superon (fermionic matter) and  
subsequently ignites the Big Bang and the inflation of the present universe.   
$L_{\rm NLSUSYGR}(w)$  may be regarded as a realization of the Mach's principle 
on new space-time, for ${e^{a}}_{\mu}$ and ${t^{a}}_{\mu}$ 
contribute equally to the unified vierbein ${w^{a}}_{\mu}$, i.e. to the whole geometry of space-time. \par
We have shown that the cosmological constant in $L_{\rm NLSUSYGR}(w)$  
is the origin of everything (all matter) and that the the vacua of SGM  
created by the Big Decay of $N$-extended NLSUSY GR action 
possess rich structures promissing for the unified description of nature.  \par 
In fact, we have shown explicitly by using NL/L SUSY relations that $N=2$ LSUSY theory of the realistic $U(1)$ gauge theory  
appears as the physical field configurations on the vacuum of $N=2$ NLSUSY GR theory on Minkowski tangent space-time, 
which gives new insights into the origin of mass, the dark matter and the dark energy of the universe.  
The cosmological implications of the composite SGM scenario deserve further studies.   \par
Interestingly the physical particle states of $N=2$ SUSY QED as a whole 
look  the similar structure to the lepton and the Higgs sector of the SM with the local $U(1)$ 
and the global $SU(2)$ \cite{STT2}. 
The neutral scalr field  $\rho (\sim m_{\nu})$ of the radial mode may be a candidate of the dark matter, 
provided the $N=2$ LSUSY QED structure is preserved in the realistic large N SUSY GUT model.  
(Note that $\omega$ in (\ref{VACb}) is a NG boson and disappears provided the corresponding local gauge symmetry 
is introduced as in the SM.) 
We anticipate that the physical cosequences obtained in $d=2$ hold in $d=4$ as well, 
for the both have the similar on-shell helicity states of $N=2$ supermultiplet 
though scalar fields and off-shell (auxiliary field) structures are modified (extended) and 
though the similar investigations in $d = 4$ are urgent for the realistic model building based upon SUSY. \par  
Further investigations on the spontaneous symmetry breaking for $N \geq 2$ SUSY remains to be studied 
for the realistic model buildings.  
The extension to large $N$, especially to $N = 5$ 
is important for {\it superon\ quintet\ (SQ) hypothesis} of SGM scenario.  
SQ model \cite{KS1} predicts a (heavy?) lepton state $L^{2+}$ with double charges. 
$N = 4$ case may shed new light on the mathematical roles of LSUSY in 
the anomaly free non-trivial $d=4$ field theory.     \par
We can anticipate that any $N$-extended LSUSY (broken) model can be related to $N$-extended NLSUSY model 
via the NL/L SUSY relation containing  the SUSY invariant relations with the universal forms  $\sim (\psi)^{n} \vert w \vert$ 
as shown in (\ref{SSUSYinv1}) and (\ref{SSUSYinv}),
which is favourable to the composite SGM viewpoint \cite{KS1,KS3} of the NLSUSY GR theory.  \par
Also NLSUSY GR with extra space-time dimensions equipped with the Big Decay is an interesting problem,       
which can give the framework for describing all observed particles as elementary {\it \`{a} la} Kaluza-Klein. \par
Linearizing $N=2$ SGM action $L_{\rm SGM}(e,\psi)$ on curved space-time, 
which elucidates the topological structure of space-time \cite{SK}, is a challenge. 
The corresponding NL/L SUSY relation will give the supergravity (SUGRA) \cite{FNF,DZ} 
analogue with the vacuum breaking SUSY spontaneously.  \par
The physical and mathematical meanings of the black hole as a singularity of space-time and 
the role of the equivalence principle are to be studied in detail in NLSUSY GR and SGM scenario .  \par   
Finally we just mention that NLSUSY GR and the subsequent SGM scenario 
for the spin ${3 \over 2}$ NG fermion \cite{ST3,BAAK} is in the same scope. 
\section{Acknowledgements}
The author (K.S.) would like to thank the organizers, especially Professor V. Dobrev and Professor D. Luest 
for inviting him to the EU RTN workshop and for the worm hospitality during the workshop.
\newpage
%


\end{document}